\documentclass[prl,twocolumn,unsortedaddress,amsmath,nofootinbib,floatfix]{revtex4-1}
\usepackage{amssymb}
\usepackage{amsmath}
\usepackage{graphicx}
\usepackage{amsbsy}
\usepackage{color}

\begin{document}
\title{The harmonics suppression effect of the quasi-periodic undulator in SASE free-electron-laser}

\author{Wu Ai-Lin}
\email{wuailing@mail.ustc.edu.cn}
\affiliation{National Synchrotron Radiation Laboratory, University of Science and Technology of China, Hefei, Anhui, 230029, China}

\author{Jia Qi-ka}
\affiliation{National Synchrotron Radiation Laboratory, University of Science and Technology of China, Hefei, Anhui, 230029, China}

\author{Li He-ting}
\affiliation{National Synchrotron Radiation Laboratory, University of Science and Technology of China, Hefei, Anhui, 230029, China}

\date{\today}

\begin{abstract}
In this paper, the harmonics suppression effect of QPUs in SASE FEL is investigated. The numerical results show that the harmonics power is reduced by using QPUs, but the fundamental radiation power also has a marked decrease as the saturation length gets very long. The cases of employing QPUs as a parts of undulators are studied. The calculations show that if the fraction of QPUs and their offgap are appropriate in an undulator system, the harmonics radiation could be suppressed remarkably, meanwhile the fundamental saturation length increases not too much.
\end{abstract}


\maketitle
\section{Introduction}

The radiation spectrum of the conventional undulator which has periodic magnetic structure is mixed with the fundamental and high order harmonics. For some accelerator users¡¯ experiments, the high harmonics suppression is very important, thus the first quasi-periodic undulator (QPU) is proposed by Hashimoto and Sasaki [1], in which the magnets are arrayed in the way of the Fibonacci sequence with two irrationally different interpole distances. The typical Halbach QPU was built and tested in ESRF [2] with the H-magnets in quasi-periodic location are displaced vertically. Recently two novel schemes of QPU are proposed and the radiation spectral fluxes demonstrate that the new schemes could greatly suppress the high-order harmonic radiation compared with the current QPU [3].

Ordinarily, the QPUs are used as insertion devices in accelerators. Therefore only the spontaneous radiation spectrum is studied in above research. Actually, the pure monochromatic light is also welcomed by FEL users. In this paper, the suppression effect of a QPU in Self-Amplified Spontaneous Emission free-electron laser (SASE FEL) is intensively studied, and compared with that of standard undulator. The numerical simulation of the undulator magnetic fields, spontaneous radiation spectrum and radiation power of SASE FEL have been carried out using the RADIA code [4], the SPECTRA code [5] and GENESIS [6], respectively.

\section{The spontaneous radiation spectrum of a QPU}

The conventional QPU structure is selected here. The magnetic structure of the QPU is determined as the Fibonacci sequence. Figure 1 shows the QPU configuration of QPU and the main parameters of the undulator are summarized in Table 1. The main parameters are based on the preliminary design research of the VUV-FEL in NSRL. The electron energy, peak current, energy spread and emittance are assumed to be 405.2MeV, 500A, 0.01\% and 1.9mm-mrad, respectively.
\begin{table}[!hbp]
\caption{\label{Table 1}Main Parameters of undulator.}
\begin{tabular}{cc}
\hline
  period length/mm & 32 \\
  period number & 50 \\
  gap/mm & 10 \\
  remanence of magnet blocks/T & 1.21 \\
  height of magnetic blocks/mm & 52 \\
  width of magnetic blocks/mm & 8 \\
  offgap \(\delta\)/mm & 3/5 \\
\hline
\end{tabular}
\end{table}
\begin{figure}[!ht]
\begin{center}
\includegraphics[width=7.3cm]{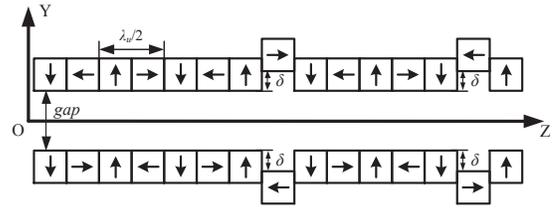}
\caption{\label{fig1} The QPU configuration.}
\end{center}
\end{figure}
The angular spectral flux densities on axis (10m from the source) from different undulators are calculated and the results are presented in Figure 2. It can be found that the 1$^{st}$ radiation of two QPUs (offgap=3mm and offgap=5mm) is lower than that of the normal case. But the advantage is that the 3$^{rd}$ and 5$^{th}$ harmonic radiation of QPU (offgap=3mm) is suppressed about 22.6\% and 13.4\% respectively. For the QPU (offgap=5mm), the suppression radio is 24.5\% and 20.8\%. As the offgap is chosen larger, the high-order harmonics will be suppressed much more efficiently.
\begin{figure}[!ht]
\begin{center}
\includegraphics[width=5cm]{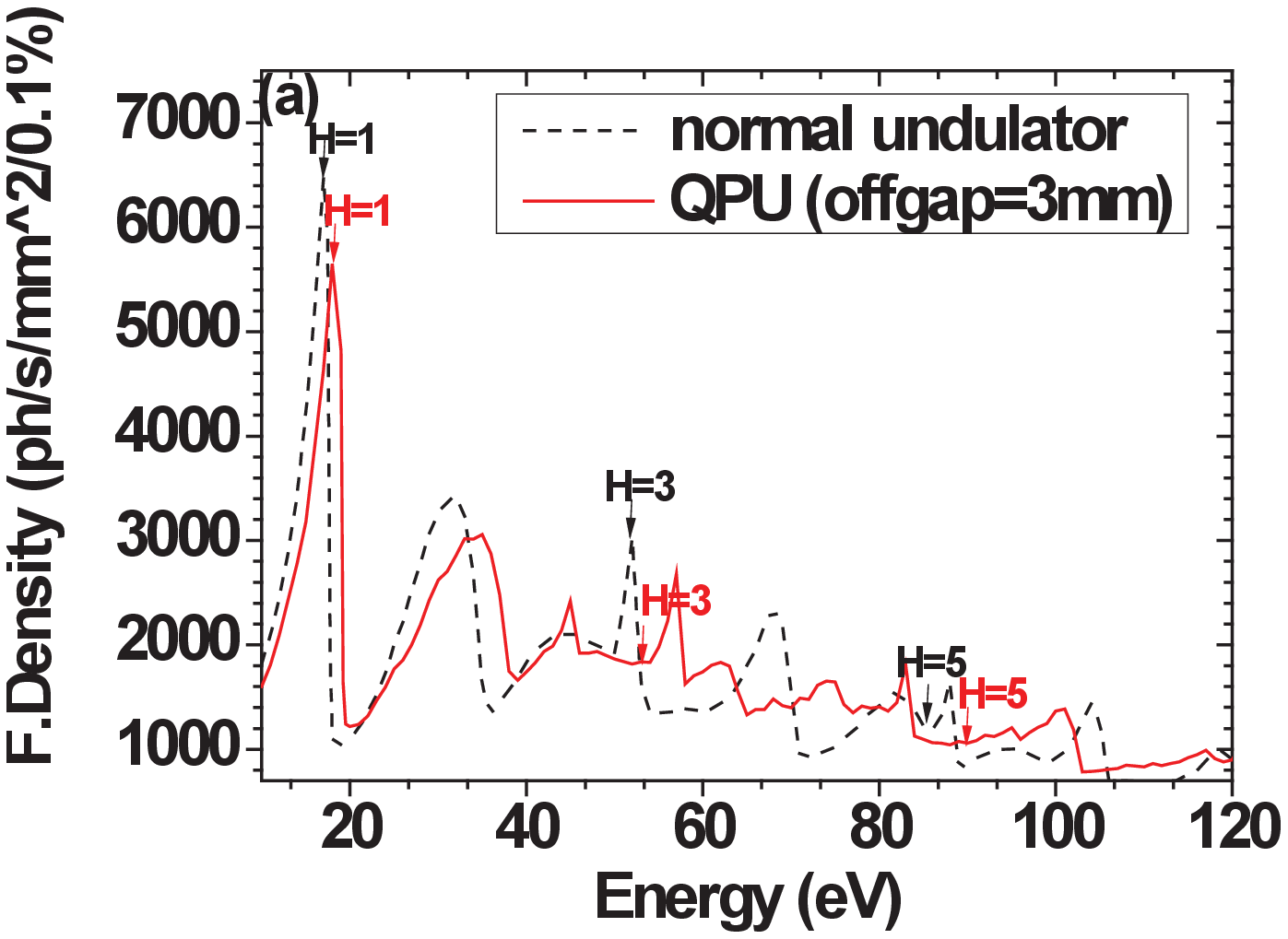}
\includegraphics[width=5cm]{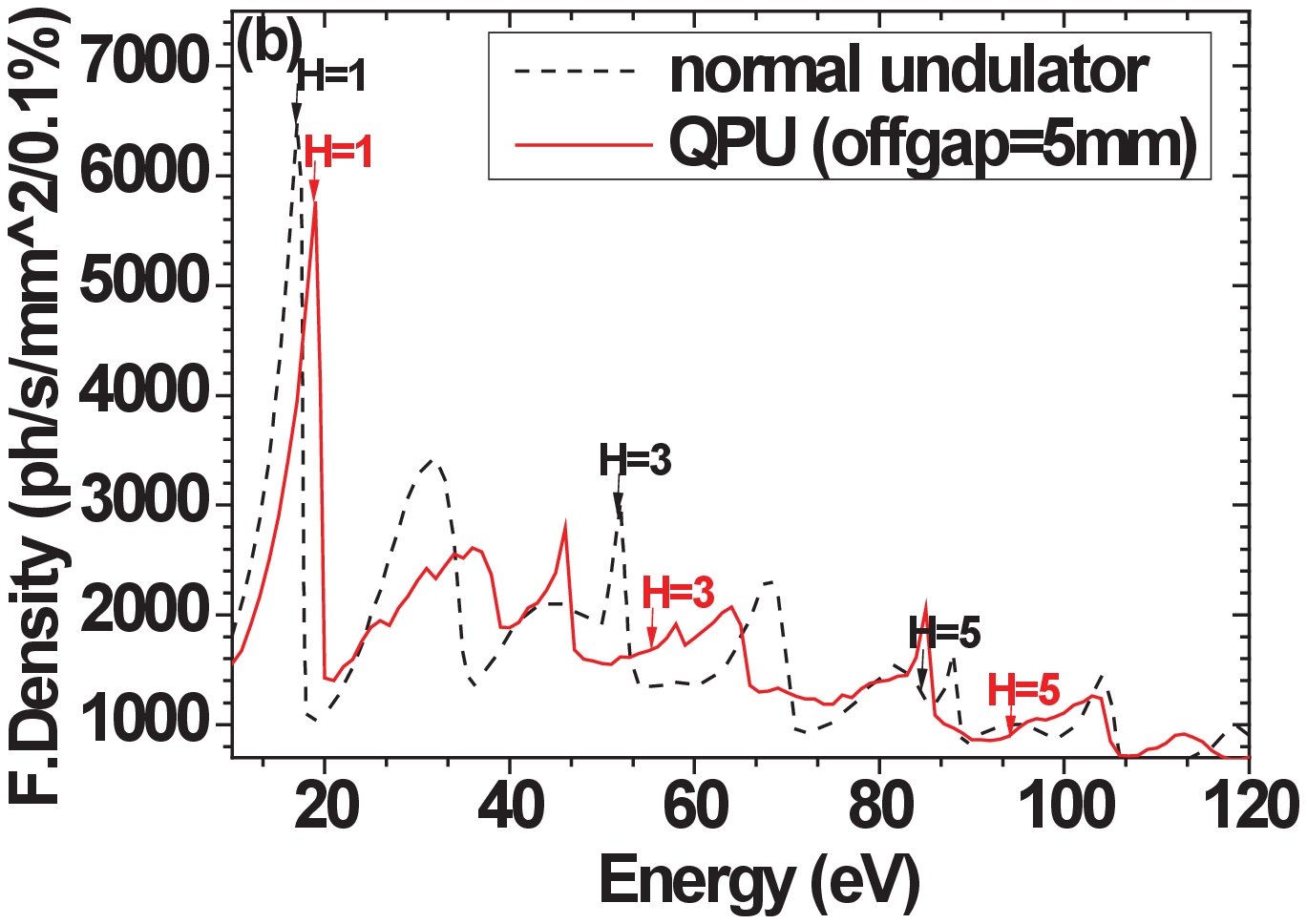}
\caption{\label{fig2} The angular spectral flux densities of the standard undualtor compared with that of the QPU: (a) offgap=3mm and (b) offgap=5mm. The 1$^{st}$, 3$^{rd}$ and 5$^{th}$ harmonic positions are marked with arrows (the red ones for the QPU).}
\end{center}
\end{figure}
\section{The radiation power of SASE FEL}

From the spontaneous radiation spectrum above, the energy of fundamental radiation becomes larger with the increases of offgap. Similarly, the results of a numerical scan indicate that the resonant wavelength of SASE FEL will decrease if we choose a big value of offgap of the QPUs. Therefore, the calculation of SASE FEL with different undulators must adopt a specific resonant wavelength in the following section. When the offgap rises from 3mm to 5mm, the resonant wavelength reduces from 65.6nm to 64.4nm.

In this paper, the SASE FEL scheme employs ten undulator sections. Firstly, the SASE FEL process based on ten normal undulators is simulated. The resonant wavelength is 70nm. The saturated power of the 1$^{st}$ radiation is about 313.5MW while the power of the 3$^{rd}$ and 5$^{th}$ harmonics is 7.2MW and 46.3kW, respectively. The simulation based on ten QPUs with different offgap is also done, but the results show that the SASE FEL could not start up as the saturation length gets too long. Needs to be pointed out is that the position of the power given in this paper is the end of entire undulators system and the total length of the undulators system is fixed.

In order to shorten the saturation length, guarantee a certain power of the 1$^{st}$ radiation while keep harmonics radiation suppressed, we considered a combined undulators system in SASE FEL. Using several normal undulators in the front to start up the SASE process and several QPUs followed to suppress the radiation of the 3$^{rd}$ and 5$^{th}$ harmonics. Corresponding radiation spectrum of this structure is calculated and analyzed. The saturated power of 1$^{st}$ radiation grows to about 192MW when using two normal undulators and eight QPUs with the offgap of 3mm. However, unlike spontaneous radiation spectrum of QPU, the growth ratios of 3$^{rd}$ and 5$^{th}$ harmonics power is about 5\% and 0.8\% compared with the SASE FEL using ten normal undulators above. Although the power of 1$^{st}$ radiation in this structure is considerable, it is not able to revitalize to suppress the high-order harmonics radiation. Figure 3(a) presents the radiation power as a function of the length of undulators. The dashed lines and solid lines stand for the SASE FEL using ten normal undulators and two normal undulators and eight QPUs with the offgap of 5mm, respectively. It can be found that the 3$^{rd}$ (red lines) and 5$^{th}$ (blue lines) harmonics radiation get a noticeable suppression, but the saturation length is still too long and the 1$^{st}$ radiation (black lines) is about 307.8MW lower than that of SASE FEL based on ten normal undualtors.
\begin{figure}[!ht]
\begin{center}
\includegraphics[width=5cm]{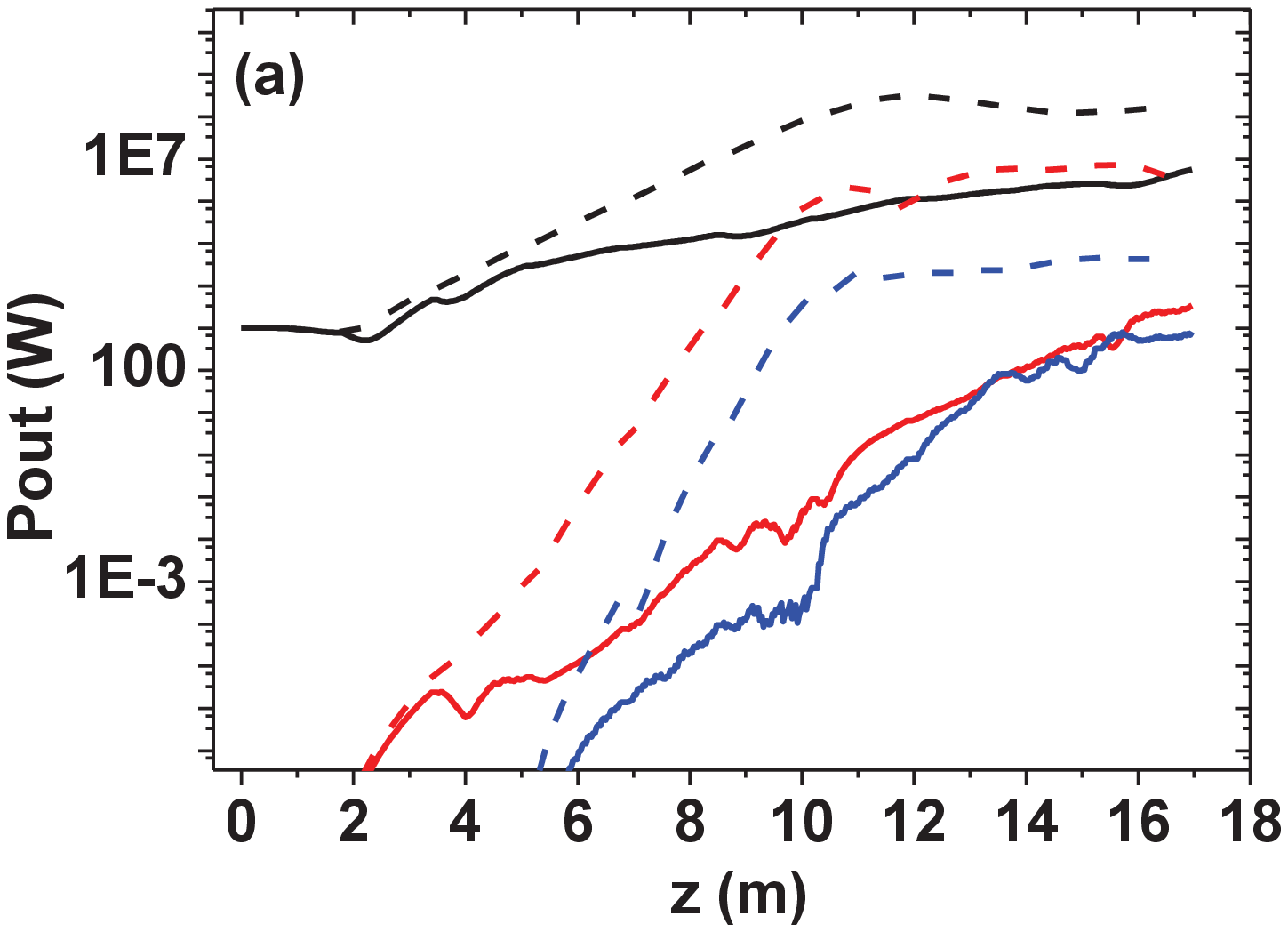}
\includegraphics[width=5cm]{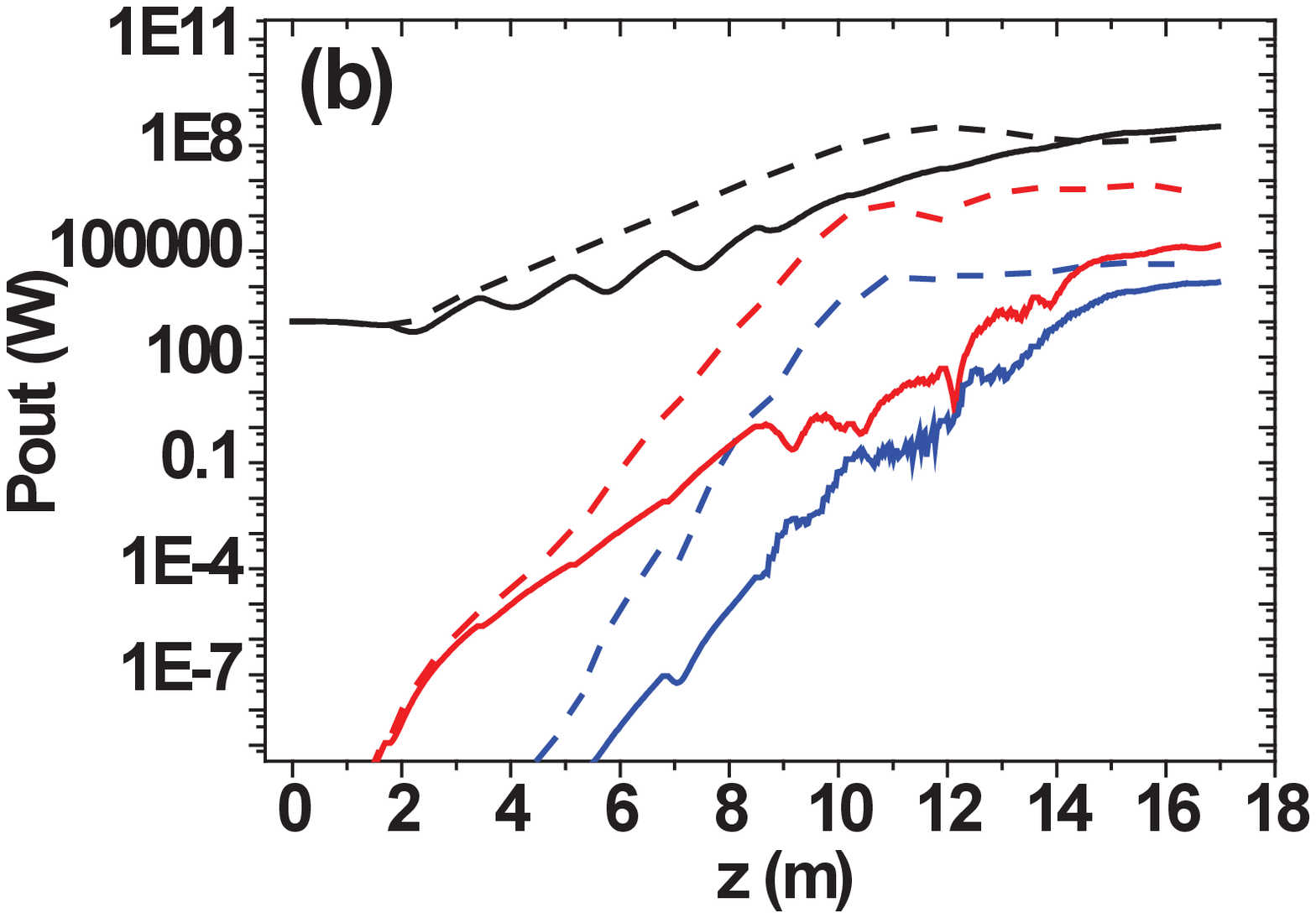}
\caption{\label{fig3} The 1$^{st}$ (black lines), 3$^{rd}$ (red lines) and 5$^{th}$ (blue lines) radiation power as a function of the length of undulators. The dashed lines stand for the FEL scheme employs ten normal undulators; the solid lines stand for the FEL scheme employs: (a) two normal undulators and eight QPUs whose offgap is 5mm and (b) five normal undulators and five QPUs whose offgap is 5mm. }
\end{center}
\end{figure}
In SASE FEL based on five normal undulators and five QPUs, if the offgap of QPUs is 3mm, the saturated power of 1$^{st}$ radiation is about 198.8MW, however, the suppression effect of high-order harmonics is still unfavorable. For the sake of comparison, the radiation power of SASE FEL which includes ten normal undulators (dashed lines) or five normal undulators and five QPUs which the offgap is 5mm (solid lines) are shown in Figure 3(b).The black lines, red lines and blue lines stand for the power of 1$^{st}$, 3$^{rd}$ and 5$^{th}$ radiation, respectively. Although the SASE FEL using five normal undulators and five QPUs could not achieve saturation, the power of 1$^{st}$ radiation is 333.8MW, even litter higher than that of conventional scheme. Moreover, the saturated power of 3$^{rd}$ radiation is about 98\% lower than that of common SASE FEL. The saturated power of 5$^{th}$ radiation is not as low as the 3rd radiation, but still much (72\%) lower than that of SASE FEL using ten normal undulators. Even at the original saturated point, the suppressive effect of high-order harmonics is also especially attractive.

\section{Conclusions}
In conclusion, we have investigated the harmonics suppressive effect of a QPU in SASE FEL. The simulation results from RADIA and GENESIS indicate that the harmonics radiation could be reduced when the normal undualtors are replaced by the QPUs. However, the fundamental radiation power is also very low since the saturation length of SASE FEL becomes quite long. Then we study the SASE FEL process in which the QPUs replace parts of the undualtor system. We find that for an appropriate combination of the normal undulators and QPUs with a proper value of offgap, the harmonics radiation could be suppressed obviously. Especially the power of 3$^{rd}$ radiation is about 98\% lower than that of SASE FEL with the normal undulators, at the same time the saturation length increases not too much.

\bibliographystyle{apsrev}

\end{document}